\documentclass[twocolumn,floatfix]{revtex4}%%%
\usepackage{epsf,graphicx}
\usepackage{amsmath,amssymb,amsfonts}
\newcommand{\nvec}[1]{\stackrel{\rightarrow}{#1}}
\newcommand{\goo}{\,\raisebox{-.5ex}{$\stackrel{>}{\scriptstyle\sim}$}\,}
\newcommand{\stdef} {\stackrel{\mbox{\scriptsize def}}{=}}
\begin{document}
\title{Heat can flow from cold to hot in
\\ Microcanonical Thermodynamics of finite systems\\
and the microscopic origin of phase transitions} \today
\author{D.H.E. Gross}
\affiliation{Hahn-Meitner Institute and Freie Universit{\"a}t Berlin,\\
Fachbereich Physik.\\ Glienickerstr. 100\\ 14109 Berlin, Germany}
\email{gross@hmi.de} \homepage{http://www.hmi.de/people/gross/ }
\begin{abstract}
Microcanonical Thermodynamics \cite{gross174} allows the
application of Statistical Mechanics to finite and even small
systems. As surface effects cannot be scaled away, one has to be
careful with the standard arguments of splitting a system into two
or bringing two systems into thermal contact with energy or
particle exchange: Not only the volume part of the entropy must be
considered. The addition of any other macroscopic constraint like
a dividing surface, or the enforcement of gradients of the
energy/particle reduce the entropy. As will be shown here, when
removing such constraint in regions of a negative heat capacity,
the system may even relax under a flow of heat {\em against the
temperature slope}. Thus Clausius formulation of the Second Law:
"Heat always flows from hot to cold" can be violated. However, the
Second Law is still satisfied and the total Boltzmann-entropy is
rising. In the final chapter the general microscopic mechanism
leading to the convexity of the microcanonical entropy at phase
separation is discussed. This is explained for the liquid--gas and
the solid--liquid transition.
\end{abstract}
\maketitle

%%%%%%%%%%%%%%%%%%%%%%%%%%%%%%%%%%%%%%%%%%%%
%% MAINMATTER
%%%%%%%%%%%%%%%%%%%%%%%%%%%%%%%%%%%%%%%%%%
\section{Introduction}
In conventional (extensive) thermodynamics the thermal equilibrium
of two systems is established by bringing them into thermal
contact which allows free energy exchange\footnote{The first two
sections discuss mainly systems that have no other macroscopic
(extensive) control parameter besides the energy. E.g. the
particle density is not changed and there are no chemical
reactions.}. Equilibrium is established when the total entropy
\begin{equation}
S_{total}=S_1+S_2\label{eq1}
\end{equation}
is maximal:
\begin{equation}
dS_{total}=dS_1+dS_2=0\label{eq2}.
\end{equation}
Under an energy flux $\Delta E_{2\to 1}$ from $2\to 1$ the total
entropy changes by
\begin{eqnarray}
\Delta S_{total}&=&(\beta_1-\beta_2)\Delta E_{2\to 1}\\
\beta&=&dS/dE=\frac{1}{T}
\end{eqnarray}
Consequently, a maximum of $S_{total}(E=E_1+E_2)$ will be approached when
\begin{equation}
 \mbox{sign}(\Delta S_{total})=\mbox{sign}(T_2-T_1)\mbox{sign}(\Delta E_{2\to 1})>0
\end{equation}
From here Clausius' first formulation of the Second Law follows: "Heat
always flows from hot to cold". Essential for this conclusion is the {\em
additivity} of $S$ under the split (eq.\ref{eq1}). Temperature is an
appropriate control parameter for extensive systems.
%%%\clearpage

\section{Stability against spontaneous energy-gradients\label{chsplit}}
\begin{figure}[h]
%%%\begin{center}%\vspace*{-3cm}
\includegraphics*[bb =84 57 383 618, angle=-180, width=6 cm,
clip=true]{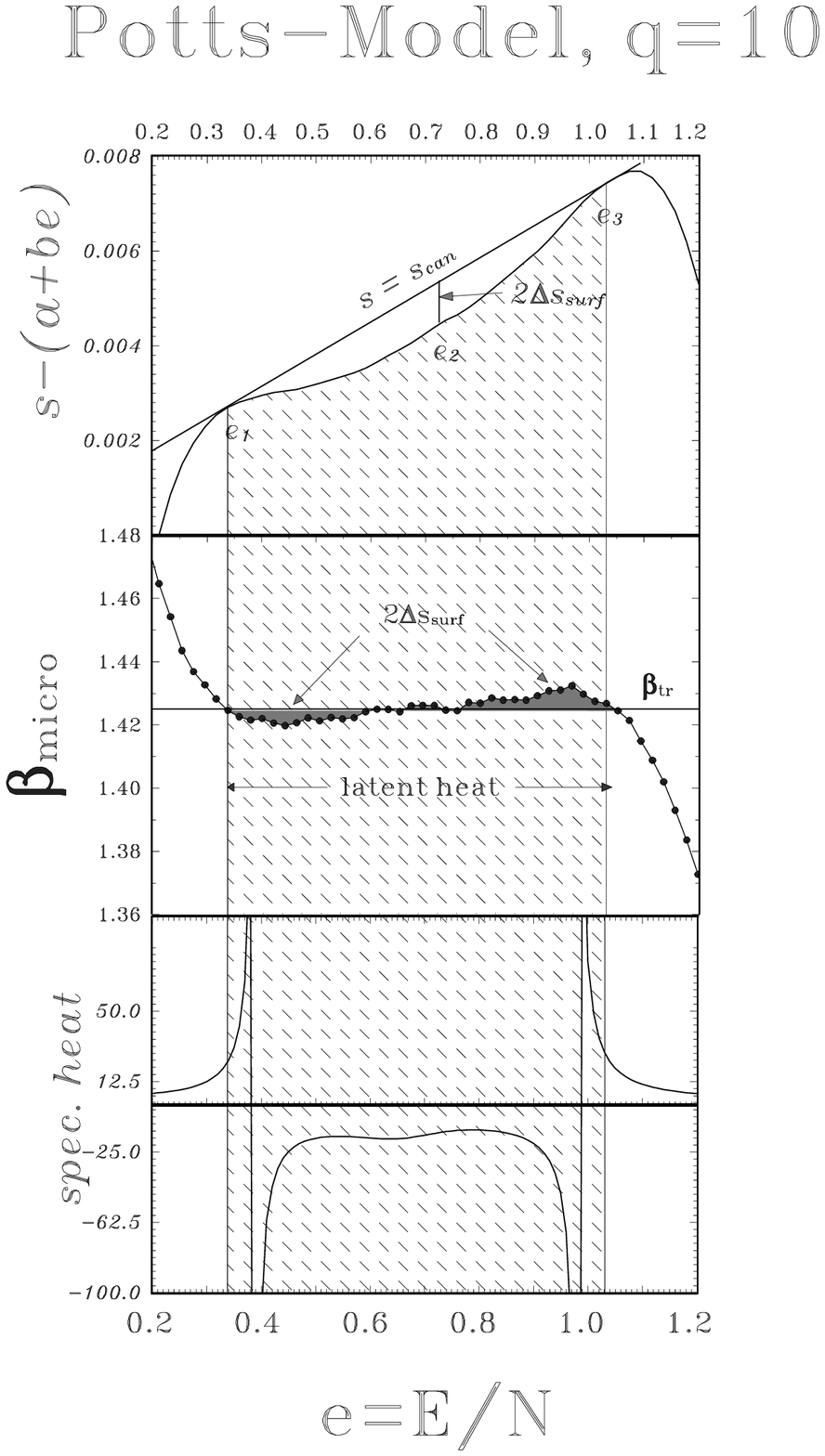}\caption{Potts model ($q=10$) on a $50*50$-lattice with
periodic boundary conditions in the region of phase separation. At the
energy $e_1$ per lattice point the system is in the pure ordered phase, at
$e_3$ in the pure disordered phase. At $e_a$ little above $e_1$ the
temperature $T_a=1/\beta$ is higher than $T_2$ and even more than $T_b$ at
$e_b$ a little below $e_3$.  At $e_a$ the system separates into bubbles of
disordered phase embedded in the ordered phase or at $e_b$ into droplets of
ordered phase within the disordered one. If we combine two equal systems:
one with the energy per lattice site $e_a=e_1+\Delta e$ and at the
temperature $T_a$, the other with the energy $e_b=e_3-\Delta e$ and at the
temperature $T_b<T_a$, and allowing for free energy exchange, then the
systems will equilibrize at energy $e_2$ with a {\em rise} of its entropy.
The temperature $T_a$ drops(cooling) and energy (heat) flows (on average)
from $b\to a$. I.e.: Heat flows from cold to hot! Clausius formulation of
the Second Law is violated. This is well known for self-gravitating
systems. However, this is not a peculiarity of only gravitating systems! It
is the generic situation at phase separations within classical
thermodynamics even for systems with short-range coupling and {\em has
nothing to do with long range interactions}.\label{figure 1}}
%%%\end{center}
\end{figure}

For a small or a very large self-gravitating system, c.f.
\cite{gross174}, additivity and extensivity of $S$, and also of
$E$, is not given. In fact this is the main reason to develop this
new and extended version of thermodynamics. At phase separation
the microcanonical caloric curve $T(E)$ is {\em backbending}. Here
the heat capacity $C=-(\frac{\partial S}{\partial
E})^2/\frac{\partial ^2S}{\partial E^2}$ becomes {\em negative},
the curvature $\frac{\partial ^2S}{\partial E^2}$ is positive, and
$S(E)$ is convex. This is the {\em generic signal of a phase
transition of first order} \cite{gross174}. The Potts-model
illuminates in a most simple example the occurrence of a
backbending caloric curve \cite{gross150}\footnote{This
backbending of $T(e)$ has nothing to do with the periodic boundary
condition used here, as claimed erroneously by Moretto et.al
\cite{moretto02}. Many realistic microcanonical calculations for
nuclear fragmentation as also for atomic-cluster fragmentation,
c.f. \cite{gross95,gross174}, show the backbending {\em without
using periodic boundaries}. These are verified by numerous
experimental data in nuclear physics c.f.
\cite{dAgostino00,borderie02} and cluster physics
\cite{schmidt01,schmidt02}. The errors of the above paper by
Moretto et. al. are commented in more detail also in
\cite{gross196,gulminelli03}}. A typical plot of $s(e)=S(E=Ne)/N$
in the region of phase separation is shown in fig(\ref{figure 1}).
Section \ref{convex} discusses the general microscopic reasons for
the convexity.

Let us split the system into two pieces $a+b$ by a dividing surface, with
half the number of particles each. The dividing surface is purely
geometrical. It exists only as long as the two pieces can be distinguished
by their different energy/particle $e_a$ and $e_b$. If the constraint on
the difference $e_b-e_a$ is fully relaxed and $e_b-e_a$ can fluctuate
freely at fixed $e_2=(e_a+e_b)/2$, the dividing surface is assumed to have
no further physical effect on the system. Constraining the
energy-difference $e_b-e_a=\Delta e$ between the two, reduces the number of
free, unconstrained degrees of freedom and {\em reduces} the entropy by
$-\Delta \Sigma$.

In the convex region (upwards concave like $y=x^2$) of $s(e)$ this becomes
especially evident: at $e=e_2 \sim e_{min}$, $e_a=e_2-\Delta e/2$ and
$e_b=e_2+\Delta e/2$, the simple algebraic sum of the individual entropies
$s_{sum}=(s_a+s_b)/2$ leads to $s_{sum}\ge s_2$. However, $e^{S_2}$ is
already the sum over {\em all} possible configurations at $e_2$ {\em
including} the split ones with $e_b-e_a=\Delta e$. Evidently, the enforced
energetic split of the system into two pieces with prescribed specific
energies $e_b>e_a$ destroys configurations with different or vanishing
$e_b-e_a$. This reduces the partition sum. It costs some additional
"surface" or constraining entropy $\Delta\Sigma>0$ to maintain the
imbalance of the energy, such that $S_2\ge S_{sum}-\Delta\Sigma $.

From this little exercise we learn that the microcanonical entropy per
particle $s(e,N,\sigma)$ of a finite system depends not only on the energy
per particle $e=E/N$ and possibly still on the particle number $N$ (for a
non-extensive system) but also on the boundary (the container), or any
other non-extensive external constraint, via $\Sigma=N\sigma$. If one
splits a given configuration like in eq.(\ref{eq1}) one must not forget the
change $-\Delta \Sigma$ in the surface entropy.

The correct entropy balance, before and after establishing the energetic
split $e_b>e_a$ of the system, is
\begin{equation}
 s_{after}-s_{before}=\frac{s_a+s_b}{2}-s_2-\Delta\sigma \le
 0 \label{balance}
\end{equation} even though the sum of the first two terms is positive.

In the inverse direction: Only by {\em relaxing} the constraint
and by allowing, on average, for an energy-flux ($\Delta e_{b\to
a}>0$) {\em opposite to $T_a-T_b>0$, against the
temperature-gradient (slope)}, but in the direction of the
energy-slope, an increase of $s_{total}\to s_2$ is obtained. Then,
on average both energies $e_a$ and $e_b$ become equal $=e_2$ and
$\Delta\sigma=0$. Or in other words, starting with an energy
difference $\Delta e=e_b-e_a\ge0$ the allowance of unconstrained
energy fluctuations around $\Delta e =0$ between the two
subsystems implies the disappearance of $\Delta\sigma$ and in
general leads to an increase of the total entropy. This is
consistent with the naive picture of an energy {\em
equilibration}. Thus Clausius' "Energy flows always from hot to
cold" is violated. Of course this shows again that {\em unlike to
extensive thermodynamics the temperature is not an appropriate
control parameter in non-extensive systems}.

In the thermodynamic limit $N\to\infty$ of a system with short-range
coupling $\Delta\Sigma\sim N^{2/3}$, $\Delta\Sigma/N=\Delta\sigma\propto
N^{-1/3}$ must go to $0$ due to van Hove's theorem.

\section{The origin of the convexity of $S(E)$ at phase separation \label{convex}}
Here I discuss the general microscopic mechanism leading to the
appearance of a convex intruder in $S(E,{\cal V},\cdots)$. This is
the signal of phase transitions of first order and of
phase-separation within the microcanonical ensemble.

I assume the system is classical and obeys the Hamiltonian:
\begin{equation}H=\sum_{i\le
j=1,N}\left\{\frac{p_i^2}{2m}\delta_{i,j}+V^{int}(\nvec{r}_i-\nvec{r}_j)\right\}
\label{hamiltonian}\end{equation} In this case the system is
controlled by energy and volume. \subsection{Liquid--gas
transition} The microcanonical sum of states or partition sum is:
\begin{widetext}
\begin{eqnarray}
W(E,N,{\cal V})&=&\frac{1}{N!(2\pi\hbar)^{3N}}\int_{{\cal
V}^N}{d^{3N}\nvec{r}_i \int{d^{3N}\nvec{p}_i{
\delta[E-\sum_i^N{\frac{\nvec{p}_i^2}{2m_i}-V^{int}\{\nvec{r}_i\}}]}}}
\nonumber\\&=&\frac{{\cal V}^N (E-E_0)^{(3N-2)/2}
\prod_1^N{m_i^{3/2}}}{N!\Gamma(3N/2) (2\pi\hbar^2)^{3N/2}}
\int_{{\cal V}^N}{\frac{d^{3N}r_i} {{\cal
V}^N}}\left(\frac{E-V^{int}\{\nvec{r}_i\}}{E-E_0}\right)^{(3N-2)/2}
\label{split}\\\nonumber\\ &=&W^{id-gas}(E-E_0,N,{\cal V})\times
W^{int}(E-E_0,N,{\cal
V})\nonumber\\\nonumber\\&=&e^{[S^{id-gas}+S^{int}]}\label{micromeg}\\&&\nonumber\\
W^{id-gas}(E,N,{\cal V})&=&\frac{{\cal
V}^NE^{(3N-2)/2}\prod_1^N{m_i^{3/2}}}{N!\Gamma(3N/2)
(2\pi\hbar^2)^{3N/2}}\\ W^{int}(E-E_0,N,{\cal V})&=&\int_{{\cal
V}^N}{\frac{d^{3N}r_i} {{\cal
V}^N}}\left(1-\frac{V^{int}\{\nvec{r}_i\}-E_0}{E-E_0}\right)^{(3N-2)/2}
\label{Win1}
 \end{eqnarray}
\end{widetext}
 ${\cal V}$ is the spatial volume. $E_0=\min
V^{int}\{\nvec{r}\}$ is the energy of the ground-state of the system.

The separation of $W(E,N,{\cal V})$ into $W^{id-gas}$ and $W^{int}$ is the
microcanonical analogue of the split of the canonical partition sum into a
kinetic part and a configuration part:
\begin{equation}
Z(T)=\frac{{\cal V}^N}{N!}\left(\frac{m
T}{2\pi\hbar^2}\right)^{3N/2}\int{\frac{d^{3N}r_i}{{\cal
V}^N}e^{-\frac{V^{int}\{\nvec{r}_i\}}{T}}}\label{canonical}
\end{equation}

In the thermodynamic limit the order parameter of the (homogeneous)
liquid-gas transition is the density. The transition is linked to a
condensation of the system towards a larger density controlled by pressure.
For a finite system we expect the analogue. However, here controlled by the
constant available system volume ${\cal V}$. If it is larger than the
eigen-volume ${\cal V}_0$ of the condensate, the system does not fill the
volume ${\cal V}$ at low $E$. $N-1$ internal coordinates are limited to
${\cal V}_0$. Only the center of mass of the droplet can move freely in
${\cal V}$. The system does not fill the $3$N-configuration space ${\cal
V}^N$. Only a stripe of the width ${\cal V}_0^{1/3}$ in $3(N-1)$ dimensions
of the total $3$-N dim space and with the length ${\cal V}^{1/3}$ in the
remaining $3$ dimensions of center of mass motion is populated. {\em The
system is non-homogeneous even though it is equilibrized and, at low
energies, internally in the single liquid phase}. It is {\em not}
characterized by an intensive homogeneous density. In fact,
$W^{int}(E-E_0,N,{\cal V})$ can be written as:
\begin{eqnarray}
W^{int}(E-E_0,N,{\cal V})&=& \left[\frac{{\cal V}(E)}{\cal V}\right]^N\le 1
\label{Win1b}\\ \left[{\cal V}(E)\right]^N&\stdef&\nonumber\\
 \lefteqn{\hspace{-3cm}\int_{{\cal
V}^N}d^{3N}r_i\left(1-\frac{V^{int}\{\nvec{r}_i\}-E_0}{E-E_0}\right)^{(3N-2)/2}}
\label{Sin2}\\\nonumber\\ S^{int}(E-E_0,N,{\cal
V})&=&N\ln\left[\frac{{\cal V}(E)}{\cal V}\right]\le0\label{Sin1}
\end{eqnarray}
In eqs.(\ref{Win1},\ref{Sin2}) the integration is only over
regions in $3$N-dim coordinate space where $V^{int}\{\nvec{r}_i\}$
is smaller than $E$, where the big brackets are positive. The
volume ${\cal V}^N(E)\le{\cal{V}}^N$ is the accessible part of the
spatial volume outside of the forbidden 3N-dim regions with
$V^{int}\{r_i\}>E$. I.e. the eigen-volume of the condensate
(droplets) at the given energy. ${\cal V}^N(E)$ has the limiting
values:
\begin{equation}
{[{\cal V}(E)]^N}:=\left\{\begin{array}{ll}{\cal V}^N&\mbox{for $E$ in the
gas phase}\nonumber\\{{\cal V}_0}^{N-1}{\cal V}&\mbox{for~~}E=E_0
\end{array}\right.\label{volume}\end{equation}
So that $W^{int}(E-E_0,N,{\cal V})$ and $S^{int}(E-E_0,N,{\cal V})$ have
the property:
\begin{eqnarray} W^{int}(E-E_0)&\le& 1,\;\Rightarrow S^{int}(E-E_0,N)
 \le0\nonumber\\
&\rightarrow& \left\{\begin{array}{ll}1 &\;\;\; \;\;\;\;\;\; \;\;\;E\gg
V^{int}\nonumber\\\left[\frac{{\cal V}_0}{\cal V}\right]^{(N-1)}&\;\;\;
\;\;\;\;\;\; \;\;\;E\to E_0\end{array}\right.
\\\\\nonumber\\ S^{int}(E-E_0)&\to
&\left\{\begin{array}{ll}0&E\gg V^{int}\nonumber\\ln\left\{[\frac{{\cal
V}_0}{\cal V}]^{N-1}\right\}< 0&E\to E_0\end{array}\right.
\label{Sin2b}\\\end{eqnarray}

 All physical details are encrypted
in $W^{int}(E-E_0,N,{\cal V})$ or $S^{int}(E-E_0,N,{\cal V})$ alias
$N\ln[{\cal V}(E)]$, c.f. eqs.(\ref{Win1b}--\ref{Sin2b}): If the energy is
high the detailed structure of $V^{int}\{\nvec{r}_i\}$ is unimportant
$W^{int}\approx 1$, $S^{int}\approx 0$. The system behaves like an ideal
gas and fills the volume ${\cal V}$. At sufficiently low energies only the
minimum of $V^{int}\{\nvec{r}_i\}$ is explored by $W^{int}(E-E_0,N,{\cal
V})$. The system is in a condensed, liquid drop (perhaps several) moving
freely inside the empty larger volume ${\cal V}$.

One can guess the general form of $N\ln[{\cal V}(E)]$: Near the groundstate
$E\goo E_0$ it must be flat $\approx(N-1)\ln[{\cal V}_0]+\ln[{\cal V}]$
because the liquid drop has some eigen-volume ${\cal V}_0$ in which {\em
each} particle can move. With rising energy $\ln[{\cal V}(E)]$ rises up to
the point ($E_{trans}$) where it is possible that the drop fissions into
two and $N\ln[{\cal V}(E)]\approx (N-2)\ln[{\cal V}_0]+2\ln[{\cal V}]$.
I.e. here is a first "jump" upwards of $S^{int}(E)$ by $\ln\{\frac{\cal
V}{{\cal V}_0}\}$. Later further such "jumps" may follow. Each of these
"jumps" induce a convex upwards bending also of the total entropy $S(E)$
(eq.\ref{micromeg}). Each is connected to a bifurcation and bimodality of
$e^{S(E)-E/T}$ and the phenomenon of {\em phase-separation}. Of course in
the conventional canonical picture for a large number of particles this is
washed out and leads to the familiar Yang-Lee singularity of the liquid to
gas phase transition. In the microcanonical ensemble this is analogue to
the phenomenon of {\em multifragmentation} in nuclear systems
\cite{gross174,gross153}. This, in contrast to the mathematical Y-L
theorem, {\em physical} microscopic explanation of the liquid to gas phase
transition shows the advantage of the microcanonical statistics on the
$3$N-coordinate space as presented here.

\subsection{Solid--liquid transition}
In contrast to the liquid phase, in the crystal phase a molecule
can only move locally within its lattice cell of the size $d^3$
instead of the whole volume ${\cal V}_0$ of the condensate. I.e.
in eq.(\ref{Sin2b}) instead we have
$S^{int}\to\ln\{[\frac{d^3}{{\cal V}_0}]^{N-1}\}$.
\subsection{Summary of section \ref{convex}}
The gas--liquid transition is linked to the transition from uniform filling
of the container volume ${\cal V}$ by the gas to the {\em smaller
eigen-volume} of the system ${\cal V}_0$ in its condensed phase where the
system is {\em non-homogeneous} (some liquid drops inside the larger {\em
empty} volume ${\cal V}$). First $N-1$, later at higher energies less and
less degrees of freedom condensate into the drop. First one, then more and
more dofs. explore the larger container volume ${\cal V}$ leading to
upwards jumps (convexities) of $S^{int}(E)$. The volume of the container
controls how close one is to the critical end-point of the transition,
where phase-separation disappears. Towards the critical end-point, i.e.
with smaller ${\cal V}$ the jumps $\ln[{\cal V}]-\ln[{\cal V}_0]$ become
smaller and smaller. In the case of the solid--liquid transition the
external volume ${\cal V}$ of the container has no influence on
$S^{int}(E)$. The entropy jumps are by $\Delta S^{int}\propto\ln{{\cal
V}_0}-\ln{d^3}$.

At the surface of the drops $V^{int}> E_0=\min V^{int}$, i.e. the
surface gives a {\em negative} contribution to $S^{int}$ in
eq.(\ref{Sin2}) and to $S$ at energies $E\goo E_0$, as was
similarly assumed in section (\ref{chsplit}) and explicitely in
eq.(\ref{balance}).
\section{Acknowledgement} I am grateful to J. M\"oller for insistent, therefore
helpful, discussions.
%%%\bibliographystyle{unsrt}%{alpha}%{plain} %{unsrt}
%%%\bibliography{gross,othbiba,othbibb,othbibcd,othbibe,othbibf,othbibg,othbibh,othbibij,othbibk,othbibl,othbibm,othbibn,othbibo,othbibp,othbibr,othbibs,othbibt,othbibuw,othbibxz}
%%%\end{document}

\end{document}